\documentclass[%
twocolumn,10pt,superscriptaddress,longbibliography,amsmath,amssymb,aps,prl,floatfix,
]{revtex4-2}

\usepackage[T1]{fontenc}

\usepackage{graphicx}
\usepackage{qcircuit}
\usepackage{mathtools}
\usepackage{amsthm}
\usepackage{tikz}
\usepackage{color}
\usepackage[colorlinks=true,linkcolor=blue,citecolor=blue, linktocpage=true,breaklinks=true]{hyperref}

\newcommand{\tr}{\mathrm{Tr}}
\newcommand{\re}{\mathrm{Re}}

\begin{document}

\title{Quasiprobability fluctuation theorem behind the spread of quantum information}

\author{Kun \surname{Zhang}}
\affiliation{School of Physics, Northwest University, Xi’an 710127, China}
\affiliation{Shaanxi Key Laboratory for Theoretical Physics Frontiers, Xi'an 710127, China}
\affiliation{Peng Huanwu Center for Fundamental Theory, Xi'an 710127, China}

\author{Jin \surname{Wang}}
\email{jin.wang.1@stonybrook.edu}
\affiliation{Center for Theoretical Interdisciplinary Sciences, Wenzhou Institute, University of Chinese Academy of Sciences, Wenzhou, Zhejiang 325001, China}
\affiliation{Department of Chemistry and of Physics and Astronomy, Stony Brook University, Stony Brook, New York 11794, USA }

\date{\today}

\begin{abstract}

\begin{center}
    \textbf{Abstract}
\end{center}

    Information spreads in time. For example, correlations dissipate when the correlated system locally couples to a third party, such as the environment. This simple but important fact forms the known quantum data-processing inequality. Here we theoretically uncover the quantum fluctuation theorem behind the quantum informational inequality. The fluctuation theorem quantitatively predicts the statistics of the underlying stochastic quantum process. To fully capture the quantum nature, the fluctuation theorem established here is extended to the quasiprobability regime. We also experimentally apply an interference-based method to measure the amplitudes composing the quasiprobability and verify our established fluctuation theorem by the IBM quantum computer.
	
\end{abstract}

\maketitle

\section{Introduction}

\begin{figure}[t]
\includegraphics[width=1\columnwidth]{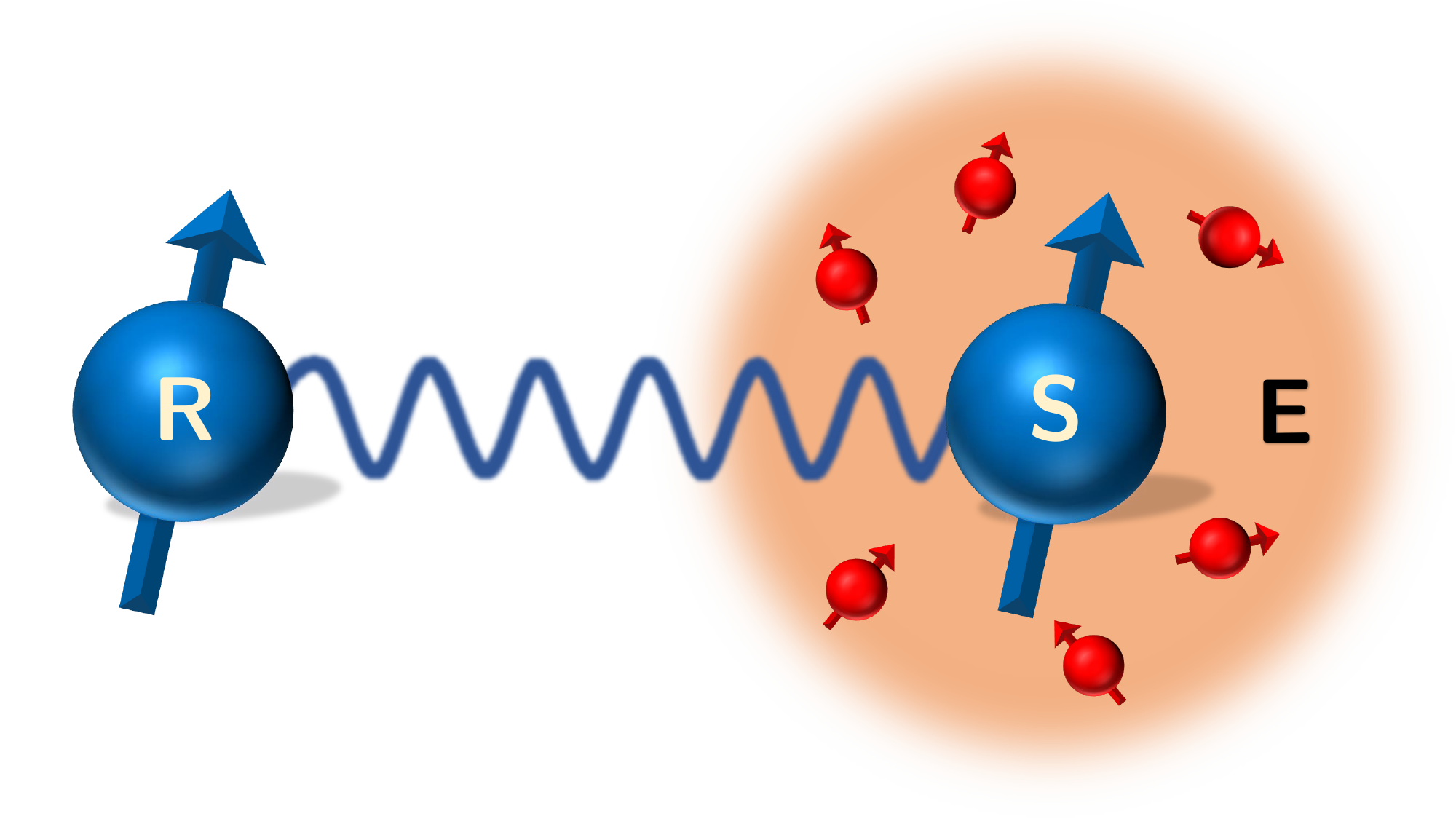}
\caption{Correlations between system and
reference. Initially the system S is correlated with the reference R. After the system locally interacts with a third party, denoted as the environment E, the correlation between S and R can only decrease if there is no initial correlations between SR and E.}
\label{fig_model}
\end{figure}

Information plays increasingly important roles in almost all branches of physics. The understanding of the dynamics of quantum information is a crucial task, which has applications in many different research subjects, such as quantum computation \cite{nielsenQuantumComputationQuantum2010}, quantum communication \cite{wildeClassicalQuantumShannon2017}, quantum thermodynamics \cite{landiIrreversibleEntropyProduction2021} as well as the information paradox in black holes \cite{almheiriEntropyHawkingRadiation2021}. In the general cases, the dynamics of quantum information can only be qualitatively described by the informational inequalities. For example, the quantum data-processing inequality states that the nonlocal information can not increase under the local physical operations \cite{schumacherQuantumDataProcessing1996}. To be specific, the amount of correlations between the system (denoted as S) and a reference (denoted as R) can only decrease if the system interacts with the environment (denoted as E) locally, with the condition that there are no initial correlations between SR and E. See Fig. \ref{fig_model} for the diagrammatic illustration. 

In a different context, since the work by Jarzynski \cite{jarzynskiNonequilibriumEqualityFree1997} and Crooks \cite{crooksNonequilibriumMeasurementsFree1998}, the qualitative description of the second law of thermodynamics (non-negativity of the entropy production) generalizes to a quantitative statement valid for arbitrary nonequilibrium processes, known as the fluctuation theorem \cite{espositoNonequilibriumFluctuationsFluctuation2009}. The inequality of the second law of thermodynamics is rephrased to an equality, which describes the distribution of entropy production. At the same time, quantum mechanics and thermodynamics are merging, and the quantum versions of fluctuation theorem are also established \cite{campisiColloquiumQuantumFluctuation2011}. Meanwhile, quantum information plays more and more important roles in the study of quantum thermodynamics, far beyond Maxwell's demon model \cite{maruyamaColloquiumPhysicsMaxwell2009,gooldRoleQuantumInformation2016,deffnerQuantumThermodynamicsIntroduction2019}. Not only the information is identified as a resource, but also the arrow of time is interpreted as the information spreading, or more specifically, the spreading of information because of coupling to the environment \cite{landiIrreversibleEntropyProduction2021}. Therefore the quantum fluctuation theorem can also be understood as a stochastic description of the correlation between the system and the environment. 

However, there is always a tension between the quantum mechanics and the classical stochastic or probabilistic description. The most famous one is Bell's theorem \cite{bellEinsteinPodolskyRosen1964}. To partially resolve the issue, quantum fluctuation theorems are commonly founded in a initial-classical and final-classical manners, in which measurements are performed at both initial and final points (the two-point measurement scheme) \cite{tasakiJarzynskiRelationsQuantum2000,campisiColloquiumQuantumFluctuation2011}. However, such scheme can only describe the case where the initial state has no correlations (between the system and the environment or within the subsystems). Therefore it is not suitable to characterize the dynamics of quantum information. Different resolutions have been proposed. However the tension between the classical and the quantum descriptions exists if the classical trajectory (described by the classical probability) is applied, such as the ones proposed in \cite{parkFluctuationTheoremArbitrary2017,abergFullyQuantumFluctuation2018,micadeiQuantumFluctuationTheorems2020,soneQuantumJarzynskiEquality2020}. On the other hand, the quasiprobability has a long history in the study of quantum dynamics \cite{dresselWeakValuesInterference2015}. Only recently, the quasiprobability plays a role in the study of quantum thermodynamics \cite{lostaglioQuantumFluctuationTheorems2018,kwonFluctuationTheoremsQuantum2019,levyQuasiprobabilityDistributionHeat2020}. 

In this study, we establish a quasiprobability fluctuation theorem which describes the dynamics of quantum information. The quasiprobability trajectory can overcome the inconsistency between the classical probabilistic description and the quantum mechanics. For the first time, we formulate the quantum data-processing inequality into the quasiprobability fluctuation theorem, which describes the statistics of the quantum information dissipation. Conceptually, our results demonstrate that the fluctuation theorem is an universal description for quantum information beyond the context of thermodynamics, if the proper quasiprobability trajectory is applied. Although quasiprobability may be hard to measure directly, there are several indirect methods, such as by the weak measurement \cite{dresselWeakValuesInterference2015}. We apply a interference-based method to measure the amplitudes on quantum computer \cite{yungerhalpernJarzynskilikeEqualityOutoftimeordered2017}, which indirectly give the quasiprobabilistic trajectory of the dynamics. Technically, we design error mitigation schemes specified for quantum circuits measuring the amplitude and the corresponding quasiprobability. We unambiguously demonstrate the quasiprobability fluctuation theorem behind the dynamics of three qubits. 

\section{Results}

\subsection{Theory}

Consider a general tripartite setup, which includes the system (S), the reference (R) and the environment (E). Initially, the system and the reference are correlated, where the initial state is denoted as $\rho_{SR}$. The amount of correlations is quantified by the mutual information $\mathcal I(S;R)_{\rho} \equiv \mathcal S(\rho_S) + \mathcal S(\rho_R) -\mathcal S(\rho_{SR})$ with the von-Neumann entropy $\mathcal S(\rho) \equiv -\tr\rho\ln(\rho)$. Both classical and quantum correlations are counted in $\mathcal I(S;R)_{\rho}$.

Assume that the system locally interacts with the environment via a unitary evolution, namely $\rho'_{SE} = U_{SE}(\rho_S\otimes \rho_E)U_{SE}^\dag$. We use the superscript prime to label variables related to the final state. The initial state of system and environment is assumed to be factorized. Therefore the correlation between the system and the environment can only increase after the evolution. On the other side, the correlation between the system and the reference is preserved in the way $\mathcal I(R;S)_\rho = \mathcal I(R;SE)_{\rho'}$. In other words, the information stored between the system and the reference spreads to the environment. The consequence is that the information between the system and the reference decreases, which gives
\begin{equation}
\label{eq:quantum_data_processing_inequality}
    \Delta \mathcal I(S;R) \geq 0,
\end{equation}
with $\Delta \mathcal I(S;R) \equiv \mathcal I(S;R)_\rho-\mathcal I(S;R)_{\rho'}$. Without any possible confusions, we simplify the notation as $\Delta \mathcal I \equiv \Delta \mathcal I(S;R)$ in the following. The above inequality is called quantum data-processing inequality \cite{schumacherQuantumDataProcessing1996}. It follows the intuition since local operations can not increase the nonlocal information. Since we assume the factorized initial state between the system and the environment, which guarantees the satisfaction of quantum data-processing inequality \cite{buscemiCompletePositivityMarkovianity2014}. The proof of classical data-processing inequality is quiet simple. Without a close analog between the classical and quantum information, the data-processing inequality also magically holds in the quantum case. The proof of quantum data-processing inequality is highly nontrivial, which usually invokes the strong subadditivity of von-Neumann entropy \cite{liebProofStrongSubadditivity1973}. The decreased mutual information $\Delta \mathcal I$ is also equal to the final state conditional mutual information between the environment and the reference (conditioned on the system), which bounds the fidelity to recover the initial state from the final state \cite{fawziQuantumConditionalMutual2015,brandaoQuantumConditionalMutual2015}. Therefore, $\Delta\mathcal I$ characterizes the degree of irretrodictability, which is rooted in thermodynamics \cite{watanabeSymmetryPhysicalLaws1955,awFluctuationTheoremsRetrodiction2021,buscemiFluctuationTheoremsBayesian2021}.

The quantum correlation between the system and the reference always flows to the environment because there is no initial correlation between the system and the environment. It resembles the heat flowing from the high temperature reservoir to the low temperature reservoir. If the thermodynamic quantity, such as heat, has a fluctuation theorem \cite{jarzynskiClassicalQuantumFluctuation2004}, do we also have a fluctuation theorem behind the spread of information? We answer this question affirmatively. 

First, we set some notations. Denote the probability distribution of the correlated state $\rho_{SR}$ as $p_l$, given by the decomposition $\rho_{SR} = \sum_l p_l\Pi_l$ with the nonlocal eigenoperator $\Pi_l = |l\rangle_{SR}\langle l|$ (projector of the eigenstate). When we look at the system and reference separately, we have the reduced density matrix $\rho_S$ and $\rho_R$, with the decompositions $\rho_R = \sum_r p_r\Pi_r$ and $\rho_S = \sum_s p_s\Pi_s$ respectively. Quantum correlation would forbid the joint distribution of $p_l$ and $p_s$ (or $p_r$), because of the noncommutativity between $\Pi_l$ and $\Pi_s$ (or $\Pi_r$). In analog to the classical stochastic mutual information (the unaveraged mutual information) \cite{parrondoThermodynamicsInformation2015}, define the stochastic quantum mutual information as $\iota(S;R)_\rho \equiv \ln p_l-\ln p_s - \ln p_r$ \cite{parkFluctuationTheoremArbitrary2017,micadeiQuantumFluctuationTheorems2020}. Similarly, the stochastic quantum mutual information of the final state $\rho'_{SR}$ is $\iota(S;R)_{\rho'} = \ln p'_{k}-\ln p'_{a} - \ln p'_{b}$, given by the decompositions $\rho'_S = \sum_a p'_a\Pi'_a$, $\rho'_R = \sum_b p'_b\Pi'_b$, and $\rho'_{SR} = \sum_k p'_k\Pi'_k$. Since the reference is untouched by the evolution $U_{SE}$, we do not expect a different distribution $p'_{b}$ from $p_r$. But the global distributions $p_l$ and $p'_{k}$ are different, since the amount of correlation between S and R decreases. The stochastic quantity behind the quantum mutual information change $\Delta\mathcal I$ is $\Delta\iota(S;R) \equiv \iota(S;R)_\rho - \iota(S;R)_{\rho'}$, with the expression
\begin{equation}
\label{eq:delta_i}
    \Delta\iota(S;R) = \ln\left(\frac{p_l}{p_sp_r}\right) - \ln\left(\frac{p'_{k}}{p'_{a}p'_{b}}\right).
\end{equation}
Without any possible confusion, we simplify the notation as $\Delta\iota \equiv \Delta\iota(S;R)$ in the following. Our starting point is to view the stochastic mutual information change $\Delta\iota$ as a stochastic entropy production. In stochastic thermodynamics, taking the average over all possible trajectories would give the average entropy production, which is always positive \cite{landiIrreversibleEntropyProduction2021}. Then the next step in our study is to identify the trajectories related to the stochastic mutual information change $\Delta\iota$.

Both classical and quantum fluctuation theorems are founded by probabilistic trajectories. However, quantum dynamics rejects the classical stochastic description, known as the temporal Bell inequalities or the Leggett-Garg inequalities \cite{emaryLeggettGargInequalities2013}. Recently, more researches have studied the quantum thermodynamics processes described by the quasiprobability, which includes quantum coherence and quantum correlation \cite{lostaglioQuantumFluctuationTheorems2018,levyQuasiprobabilityDistributionHeat2020,kwonFluctuationTheoremsQuantum2019}. Inspired from the quasiprobability in the study of quantum chaos \cite{yungerhalpernJarzynskilikeEqualityOutoftimeordered2017,yungerhalpernQuasiprobabilityOutoftimeorderedCorrelator2018}, we consider the following quasiprobabilistic trajectory
\begin{equation}
\label{eq:Q_zeta}
    \mathcal Q[\zeta] \equiv \re\tr\left(U^\dag_{SE}\Pi'_{km}\Pi'_{ab}U_{SE}\Pi_{rs}\Pi_{ln}\rho_{RSE}\right),
\end{equation}
with the stochastic variables $\zeta = \{s,r,l,n,a,b,k,m\}$. The projectors $\Pi_n$ and $\Pi'_m$ are given by the eigenstates of the environment, namely $\rho_E = \sum_n p_n\Pi_n$ and $\rho'_E = \sum_m p'_m\Pi'_m$. We adopt the simplified notations $\Pi_{r,s} \equiv \Pi_r\otimes \Pi_s$. Note that the initial state is factorized $\rho_{RSE} = \rho_{SR}\otimes \rho_E$. We only consider the real part of quasiprobability, where its imaginary part has its own interests in quantum dynamics \cite{dresselSignificanceImaginaryPart2012}.  

It is easy to see that quasiprobability $\mathcal Q[\zeta]$ is properly normalized, i.e., $\sum_\zeta \mathcal Q[\zeta] = 1$. However, it is not a valid probability because its range is not bounded between 0 and 1. The out-of-range quasiprobability is the signature of quantum interference \cite{dresselWeakValuesInterference2015}. The global operator, such as $\Pi_l$, is intertwined with the local operator, such as $\Pi_r$ and $\Pi_s$. When we marginalize the variables in terms of local or global operators, the quasiprobability returns back to a valid probability, which is exactly the probabilistic trajectories studied in quantum fluctuation theorem. For example, we have
\begin{align}
\label{eq:P[gamma]}
    \mathcal P[\gamma] \equiv & \sum_{l,k,r,b}\mathcal Q[\zeta]  \nonumber \\
    = & |\langle am|U_{SE}|sn\rangle|^2 p_sp_n,
\end{align}
with $\gamma = \{s,n,a,m\}$. Here $\mathcal P[\gamma]$ describes a probabilistic distribution of the initial and final states (in terms of the system and the environment). Here $\mathcal P[\gamma]$ is commonly applied to the study of quantum fluctuation theorem of the entropy production \cite{landiIrreversibleEntropyProduction2021}. Therefore the quasiprobabilistic trajectory can also properly describe the original quantum fluctuation theorems. For more discussions, see the Methods. 

The quasiprobability is not directly measurable. However, it is not merely a mathematical artifice. Quasiprobability describes the weak measurement, where the system is probed indirectly. More specifically, the system is weakly coupled to the measurement apparatus, then the apparatus is measured. The weak measurement scheme can also be applied to measure $\mathcal Q[\zeta]$, where the two-point weak measurement is applied. 

Firstly, we can verify that averaging the stochastic mutual information change $\Delta \iota$ over the quasiprobabilistic trajectory $\mathcal Q[\zeta]$ gives
\begin{equation}
\label{eq:averaged_delta_i}
    \langle \Delta \iota\rangle_{\mathcal Q[\zeta]} = \Delta \mathcal I.
\end{equation}
Here the bracket means the average $\langle \cdot\rangle_{\mathcal Q[\zeta]} = \sum_\zeta \mathcal Q[\zeta] (\cdot)$. Then our main result is the following quasiprobability fluctuation theorem
\begin{equation}
\label{eq:FT}
    \langle e^{-\Delta\iota}\rangle_{\mathcal Q[\zeta]} = 1.
\end{equation}
It takes the exact same form of the fluctuation theorem, while the difference is the quasiprobability applied to the trajectory. We present the derivations of Eqs. (\ref{eq:averaged_delta_i}) and (\ref{eq:FT}) in the Methods. In the Methods, we also argue that the quasiprobability trajectory is inevitable to describe the quantum information dissipation. Even if we extend the traditional two-point measurement scheme, such as proposed recently in \cite{micadeiQuantumFluctuationTheorems2020}, the classical trajectory always contradicts to the principle of quantum contextuality \cite{budroniQuantumContextuality2022}. The positivity of the averaged stochastic entities, such as the stochastic entropy production averaged over the probabilistic trajectories, simply reflect the positivity of the relative entropy \cite{landiIrreversibleEntropyProduction2021}. However, the positivity of the averaged stochastic mutual information change over the quasiprobabilistic trajectory is beyond any classical statistical inequality. 

From the quantum data-processing inequality, we know $\langle \Delta \iota\rangle_{\mathcal Q[\zeta]}\geq 0$. Therefore, combining the quasiprobability fluctuation theorem (\ref{eq:FT}), we can establish the inequality
\begin{equation}
\label{eq:quantum_jensen_inequality}
    \langle e^{-\Delta\iota}\rangle_{\mathcal Q[\zeta]}\geq e^{-\langle \Delta \iota\rangle_{\mathcal Q[\zeta]}}.
\end{equation}
It resembles the Jensen's inequality \cite{dekking2005modern}. However, Jensen's inequality would not hold if the probability has negative values. But the inequality (\ref{eq:quantum_jensen_inequality}) is guaranteed because of the quantum data-processing inequality and our established quasiprobability fluctuation theorem (\ref{eq:FT}). Even under the presence of negative values of quasiprobability, the Jensen's inequality still holds. In other words, the quasiprobability fluctuation theorem   (\ref{eq:FT}) contains constrains on the distributions of the negative quasiprobabilities, which do not break the Jensen's inequality and lead to the quantum data-processing inequality.

In the small dissipation regime, expanding to the second order of $\Delta\iota$, we get an informational fluctuation-dissipation relation
\begin{equation}
\label{eq:FD_theorem}
    \langle \left(\Delta\iota-\Delta\mathcal I\right)^2\rangle_{\mathcal Q[\zeta]} = 2\Delta\mathcal I +\mathcal O(\Delta\iota^3),
\end{equation}
where the left-hand side is the variance of the mutual information change and the right-hand side represents the degree of the information decay. One can learn more about the system-environment evolution based on the high-order distribution of $\Delta\iota$. For example, the black hole scrambling \cite{almheiriEntropyHawkingRadiation2021} may be probed by the statistics of the outgoing radiation based on the fluctuation theorem (\ref{eq:FT}). When the dissipation is large, the high-order statistics is required.

In the thermodynamic context, the fluctuation theorem regarding to the classical information dissipation firstly appeared in \cite{zengNewFluctuationTheorems}. It characterizes the extra thermodynamic cost in Maxwell's demon model. Later, the dissipative mutual information is studied in quantum cases, within the framework of two-point measurement scheme \cite{zhangConditionalEntropyProduction2021a}. The issue of the two-point measurement scheme is that the measurement performed initially would kill the quantum correlation, therefore only classical correlation is left.

When we looks closer on the fluctuation theorem in Eq. (\ref{eq:FT}), there are hidden fluctuation theorems, which have the form $\langle e^{-\Delta\iota}\rangle_{\mathcal Q[\zeta]/\mathcal P[\gamma]} = 1$. The average is done on the conditional quasiprobability $\mathcal Q[\zeta]/\mathcal P[\gamma]$ (assuming $\mathcal P[\gamma]$ is nonzero), where $\mathcal P[\gamma]$ is a valid probability given in Eq. (\ref{eq:P[gamma]}). The conditional quasiprobability $\mathcal Q[\zeta]/\mathcal P[\gamma]$ also gives the detailed fluctuation theorem. More details can be found in the Methods.

There is an ambiguity of the operator order in $\mathcal Q[\zeta]$, since operator $\Pi_l$ does not commute with $\Pi_r$ or $\Pi_s$ in general cases. However, there is no restrictions on choosing a specific order to give the quasiprobability fluctuation theorem, even for their combinations. For simplicity, here we only focus on this specific order. When choosing different operator orders, we need slightly change the definition of $\Delta\iota$. But the forms of fluctuation theorem are identical. More details can be found in the Methods and Supplementary Methods 2.

\begin{figure*}[t]
\centering
\includegraphics[width=\textwidth]{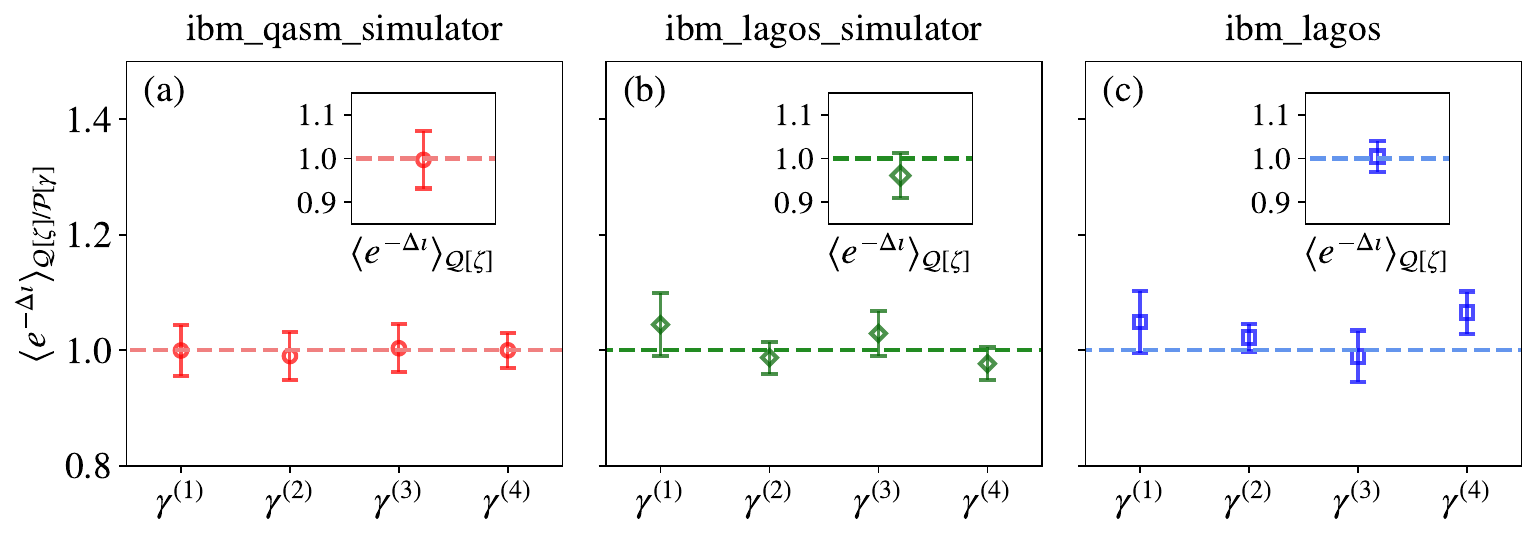}
\caption{Verification of quasiprobability fluctuation theorem on IBM quantum processor. The processors (a) {\fontfamily{qcr}\selectfont ibm\_qasm\_simulator} and (b) {\fontfamily{qcr}\selectfont ibm\_lagos\_simulator} are provided by Qiskit, which classically simulates the quantum computers without and with the noises. Results obtained from IBM quantum processor {\fontfamily{qcr}\selectfont ibm\_lagos} are presented in (c). The conditional stochastic variables $\gamma=\{s,n,a,m\}$ have the values
$\gamma^{(1)}=\{0,0,0,0\}$, $\gamma^{(2)}=\{1,0,0,1\}$, $\gamma^{(3)}=\{0,1,1,0\}$ and $\gamma^{(4)}=\{1,1,1,1\}$. The error bars are given by $10\times 8192$ shots of the circuits.}
\label{fig_ift}
\end{figure*}

\subsection{Experiment}

Quantum fluctuation theorems have been verified on many different platforms as well as different methods \cite{dornerExtractingQuantumWork2013,batalhaoExperimentalReconstructionWork2014,anExperimentalTestQuantum2015,cerisolaUsingQuantumWork2017,masuyamaInformationtoworkConversionMaxwell2018,micadeiExperimentalValidationFully2021}. Cloud quantum computers provide an unique platform to study the quantum phenomena. Quantum fluctuation theorems have also been studied on quantum computers recently \cite{solfanelliExperimentalVerificationFluctuation2021}. Quasiprobability fluctuation theorem presented above does not limit the size of the environment or types of the system-environment interaction. In the following, we consider the dynamics of three qubits and verify the corresponding fluctuation theorem through the IBM quantum computer \cite{IBM}. 

Suppose that the system, the reference and the environment are all one qubit respectively. We consider the initial correlation between the system and the reference as a full rank Bell-diagonal state, which has four Bell states as the eigenstates, i.e., $\rho_{SR} = \sum_l p_l|\psi(l)\rangle\langle\psi(l)|$ with the four Bell states $|\psi(l)\rangle$. Bell-diagonal states can be easily prepared on quantum computers with the help of two ancillary qubits \cite{pozzobomPreparingTunableBelldiagonal2019}. We randomly generate the distribution $p_l$ then perform the Bell measurements on the initial state. The environment qubit is chosen as thermal: $\rho_E = p_{n=0}|0\rangle_E\langle 0 |+p_{n=1}|1\rangle_E\langle 1|$ with $p_{n=0} = 1/(1+e^{-\beta})$ and $p_{n=1} = 1/(1+e^{\beta})$. Here $\beta$ is the effective temperature. The statistics of local states $\rho_S$, $\rho_R$ and $\rho_E$ are separately measured. 

We consider the interaction between the system qubit and the environment qubit as $U_{SE} = |0\rangle_S\langle0|\otimes 1\!\!1_E+|1\rangle_S\langle1|\otimes Y_E$ with the single-qubit gate $Y = |0\rangle\langle1|-|1\rangle\langle0|$. The two-qubit operation $U_{SE}$ has the maximal entangling power, which would disentangle the initial quantum correlation between the system and the reference. Therefore, the eigenvectors of $\rho'_{SR}$ are the same as the local eigenvectors of $\rho'_R$ and $\rho'_S$. Similarly with the initial state, we measure the statistics of the final state. The measured initial and final statistics can be found in the Methods.


In our example, the quasiprobability $\mathcal Q[\zeta]$ has the equivalent expression
\begin{equation}
\label{eq:example_Q}
   \mathcal Q[\zeta] = \langle \psi(l) n|U^\dag_{SE}|bam\rangle\langle am|U_{SE}|sn\rangle\langle rs|\psi(l)\rangle p_l p_n,
\end{equation}
with the short notation $|\psi(l)n\rangle = |\psi(l)\rangle\otimes |n\rangle$. Although sequential weak measurements can be applied to infer the value of $\mathcal Q[\zeta]$, it is still challenging for quantum computers to perform precise controls. We follow a interference-based method \cite{yungerhalpernJarzynskilikeEqualityOutoftimeordered2017} to measure the amplitudes $\langle \psi(l) n|U_{SE}|bam\rangle$, $\langle am|U_{SE}|sn\rangle$ and $\langle rs|\psi(l)\rangle$ separately. See the Methods for the experimental results. More details on the experimental setup and the corresponding quantum circuits can also be found in the Methods.

Now we can assemble together the stochastic mutual information change $\Delta\iota$ with the quasiprobabilistic trajectory $\mathcal Q[\zeta]$ to verify the fluctuation theorem. The fluctuation theorem $\langle e^{-\Delta\iota}\rangle_{\mathcal Q[\zeta]/\mathcal P[\gamma]} = 1$ is based on the conditional quasiprobability $\mathcal Q[\zeta]/\mathcal P[\gamma]$. For each variable $\gamma$ which gives a nonzero $\mathcal P[\gamma]$, we have a fluctuation relation. The experimental results are presented in Fig. \ref{fig_ift}. Results from the quantum computer are little deviated from the noisy or noiseless simulation results, which imply that the deviations are from the imperfect operations of near-term quantum computers. 

\section{Discussions}

Fluctuation theorem is one of the most important tools to study the nonequilibrium dynamics. Most quantum fluctuation theorems are founded on converting the quantum dynamics into a classical probabilistic description, in which many quantum features are lost, such as the coherence and the entanglement. Quasiprobability has a long history on studying the quantum dynamics. However, quasiprobability is rarely considered in the study of quantum fluctuation theorem. For the first time, we construct the quasiprobability fluctuation theorem behind the quantum data-processing inequality. We establish such fluctuation theorem beyond the thermodynamic regime, namely neither having any constraints on the interactions between the system and the environment, nor limited to specific initial and final states. We argue that the quasiprobability trajectory is necessary in order to correctly describe the dynamics of quantum information. The fluctuation theorem predicts the statistics of quantum information processing, such as the informational fluctuation-dissipation relation. We also test our quasiprobability fluctuation theorem on the state-of-art quantum computers. Technically, we design quantum circuits on measuring the quasiprobability trajectory of three qubits. Our study generalizes the subject of stochastic thermodynamics into the stochastic quantum information, which provides novel insights on the dynamics of quantum information. 

Independently, the fluctuation theorem of the monotonicity of quantum relative entropy was proposed in \cite{kwonFluctuationTheoremsQuantum2019}. Quasiprobabilistic trajectories were also applied. It would be interesting to clarify the relations between their fluctuation theorems and our work.

Our informational fluctuation theorem suggests hidden constraints on the negative quasiprobability distributions, which is required by the quantum data-processing inequality. In other words, a ``quantum version of Jensen's inequality'' is required to derive the quantum data-processing inequality from our informational fluctuation theorem. Statistical analysis can be carried out on the traditional fluctuation theorem \cite{merhavStatisticalPropertiesEntropy2010}. It would be interesting to explore along the same line for the quasiprobability fluctuation theorem in future. When the quantum data-processing inequality breaks, whether the quasiprobability fluctuation theorem can capture the anomalous flow of quantum information is an open question. It is related to the non-Markovian dynamics. Attempts have been made recently \cite{huangFluctuationTheoremsMultitime2022}. 

\section{Methods}

\subsection{Averaged mutual information change}

The quasiprobabilistic trajectory $\mathcal Q[\zeta]$ reduces to the probabilistic trajectory $\mathcal P[\gamma]$ after marginalizing the stochastic variables $\{l,k,r,b\}$, shown in Eq. (\ref{eq:P[gamma]}). The trajectory $\mathcal P[\gamma]$ appears in the ordinary quantum fluctuation theorem within the two-point measurement scheme \cite{landiIrreversibleEntropyProduction2021}. Consider the stochastic entropy production of the system, given by $\sigma_S \equiv \ln p_s + \ln p_n - \ln p'_{a} - \ln p'_{m}$. One can easily find that the averaged entropy production over the trajectory $\mathcal P[\gamma]$ is
\begin{equation}
\label{eq:average_s}
    \langle\sigma_S\rangle_{\mathcal P[\gamma]} = \mathcal I(S;E)_{\rho'}.
\end{equation}
Different stochastic quantity can give the different averaged quantity. Here we set the stochastic entropy production $\sigma_S$ which gives the mutual information between the system and the environment established on the final state. 

The quasiprobabilistic trajectory $\mathcal Q[\zeta]$ has another way to reduce to the probabilistic trajectory. Marginalizing the stochastic variables $\{s,r,a,b\}$ gives
\begin{align}
\label{eq:P'[tau]}
    \bar{\mathcal P}[\tau]\equiv& \sum_{s,r,a,b}\mathcal Q[\zeta] \nonumber \\
    = & |\langle km|U_{SE}|ln\rangle|^2 p_lp_n,
\end{align}
with $\tau = \{l,n,k,m\}$. We can consider the stochastic entropy production of the combined system and reference, given by $\sigma_{SR} = \ln p_l + \ln p_n - \ln p'_{k} - \ln p'_{m}$. Then the corresponding averaged entropy production over the trajectory $\bar{\mathcal P}[\tau]$ is 
\begin{equation}
\label{eq:average_sr}
    \langle\sigma_{SR}\rangle_{\bar{\mathcal P}[\tau]} = \mathcal I(SR;E)_{\rho'}.
\end{equation}
Physically, the stochastic entropy production $\sigma_S$ ($\sigma_{SR}$) characterizes the correlation between the system (system plus the reference) and the environment as irreversibility. 

The stochastic mutual information change $\Delta \iota$ given in Eq. (\ref{eq:delta_i}) is also the difference between $\sigma_{SR}$ and $\sigma_{S}$, namely $\Delta \iota = \sigma_{SR} - \sigma_{S}$ (with $r=b$). Taking the average over the quasiprobabilistic trajectory $\mathcal Q[\zeta]$, we have
\begin{align}
    \langle\Delta\iota\rangle_{\mathcal Q[\zeta]} = &  \sum_{\tau}\sum_{s,r,a,b}\mathcal Q[\zeta]\sigma_{SR} -  \sum_{\gamma}\sum_{l,r,k,b}\mathcal Q[\zeta]\sigma_{S} \nonumber \\
    = & \sum_{\tau} \bar{\mathcal P}[\tau]\sigma_{SR} - \sum_{\gamma} \mathcal P[\gamma]\sigma_{S},
\end{align}
where the marginalizing relation (\ref{eq:P'[tau]}) have been applied to the second line. Based on the averaged quantities in Eqs. (\ref{eq:average_s}) and (\ref{eq:average_sr}), we find $\langle\Delta\iota\rangle_{\mathcal Q[\zeta]} = \mathcal I(SR;E)_{\rho'} - \mathcal I(S;E)_{\rho'}$, which is also equal to $\langle\Delta\iota\rangle_{\mathcal Q[\zeta]} = \mathcal I(S;E)_{\rho} - \mathcal I(S;E)_{\rho'}$, since the initial correlation is preserved given by $\mathcal I(S;E)_{\rho} = \mathcal I(SR;E)_{\rho'}$. Then we prove that the averaged stochastic mutual information change $\Delta\iota$ equals to $\Delta \mathcal I$.

\subsection{Proof of the quasiprobability fluctuation theorem}

The integral fluctuation theorem relies on the normalization of a corresponding retrodictive trajectory (also commonly known as the backward process) \cite{awFluctuationTheoremsRetrodiction2021,buscemiFluctuationTheoremsBayesian2021}. Therefore the choice of retrodictive process is subjective. Similar in our quasiprobability fluctuation theorem, the integral version is established on the normalization of the quantum retrodictive trajectory, which is given by the conditional quasiprobability (quasiprobability conditioned on a probability). 

Corresponding to the forward quasiprobabilistic trajectory $\mathcal Q[\zeta]$ defined in Eq. (\ref{eq:Q_zeta}), we consider the retrodictive quasiprobabilistic trajectory
\begin{equation}
    \tilde{\mathcal Q}[\zeta] \equiv \re\tr\left(U_{SE}\Pi_{ln}\Pi_{rs}U^\dag_{SE}\Pi'_{ab}\Pi'_{km}\rho'_{SR}\otimes\rho'_{E}\right),
\end{equation}
where the initial state is set as the factorized final state of the system plus the reference and the environment. Marginalizing the variables $\{l,r,k,b\}$ gives the probability
\begin{align}
    \tilde{\mathcal P}[\gamma]\equiv & \sum_{l,r,k,b}\tilde{\mathcal Q}[\zeta] \nonumber \\
    = & |\langle sn|U_{SE}^\dag|am\rangle|^2 p'_{a}p'_{m},
\end{align}
where $p'_{a}$ and $p'_{m}$ are the eigenvalue distributions of $\rho'_{S}$ and $\rho'_{E}$ respectively.  

After some algebra, we can establish the relation
\begin{equation}
\label{eq:Q_P_realtion}
    \frac{\mathcal Q[\zeta]}{\mathcal P[\gamma]}e^{-\Delta\iota} = \frac{\tilde{\mathcal Q}[\zeta]}{\tilde{\mathcal P}[\gamma]}.
\end{equation}
The ratio between the conditional forward quasiprobabilistic trajectory $\mathcal Q[\zeta]/\mathcal P[\gamma]$ and the conditional retrodictive quasiprobabilistic trajectory $\tilde{\mathcal Q}[\zeta]/\tilde{\mathcal P}[\gamma]$ is given by $e^{-\Delta\iota}$. Their ratio is exponential scaled with the stochastic mutual information change $\Delta\iota$, which is in the form of the detailed fluctuation theorem. Here the probabilistic distribution has been generalized to the conditional quasiprobability distribution. Detailed proof of Eq. (\ref{eq:Q_P_realtion}) can be found in the Supplementary Methods 1.

The conditional quasiprobabilistic trajectories are properly normalized 
\begin{equation}
    \sum_{\zeta/\gamma} \frac{\mathcal Q[\zeta]}{\mathcal P[\gamma]} = \sum_{\zeta/\gamma}\frac{\tilde{\mathcal Q}[\zeta]}{\tilde{\mathcal P}[\gamma]} = 1.
\end{equation}
Then we have the quasiprobability fluctuation theorem given by the conditional quasiprobabilistic trajectory
\begin{equation}
    \langle e^{-\Delta\iota}\rangle_{\mathcal Q[\zeta]/\mathcal P[\gamma]} = \sum_{\zeta/\gamma}\frac{\tilde{\mathcal Q}[\zeta]}{\tilde{\mathcal P}[\gamma]} = 1.
\end{equation}
Moreover, the unconditional quasiprobabilistic trajectory also gives the integral fluctuation theorem
\begin{equation}
    \langle e^{-\Delta\iota}\rangle_{\mathcal Q[\zeta]} = \sum_{\gamma} \mathcal P[\gamma] \sum_{\zeta/\gamma}\frac{\tilde{\mathcal Q}[\zeta]}{\tilde{\mathcal P}[\gamma]} = 1,
\end{equation}
which is guaranteed by the normalization of $\mathcal P[\gamma]$. With the help of Eq. (\ref{eq:Q_P_realtion}), the quasiprobability fluctuation theorem simply comes from the normalization of retrodictive quasiprobability trajectory. However, we can also directly prove the quasiprobability fluctuation theorem without the definition of retrodictive quasiprobability trajectory. See the Supplementary Methods 1.

Different ordering of projectors in $\mathcal Q[\zeta]$ does not jeopardize the validation of the quasiprobability fluctuation theorem, as long as the ordering of projectors in $\tilde{\mathcal Q}[\zeta]$ is correspondingly changed and the relation (\ref{eq:Q_P_realtion}) holds. Besides the averaged mutual information change does not require specific orderings of projectors in $\mathcal Q[\zeta]$, since the marginalized relation, such as Eq. (\ref{eq:P'[tau]}), eliminate the uncommuted operators. 

In addition to the quasiprobability fluctuation theorem revealed above, the quasiprobabilistic trajectory $\mathcal Q[\zeta]$ also smoothly generalize the quantum fluctuation theorem of the entropy productions $\sigma_S$ and $\sigma_{SR}$. Specifically, we have
\begin{equation}
\label{eq:sigma_s_sr}
    \langle e^{-\sigma_S}\rangle_{\mathcal Q[\zeta]} = \langle e^{-\sigma_{SR}}\rangle_{\mathcal Q[\zeta]} = 1.
\end{equation}
We can simply have the proof by applying the marginalizing rules shown in Eq. (\ref{eq:P'[tau]}). Then the above quasiprobability fluctuation theorems reduce to the well-known quantum fluctuation theorems given by the two-point measurement trajectories \cite{landiIrreversibleEntropyProduction2021}.

\subsection{Probability vs Quasiprobability trajectories}

Quantum fluctuation theorems are commonly established based on the two-point measurement scheme \cite{tasakiJarzynskiRelationsQuantum2000,campisiColloquiumQuantumFluctuation2011}. Then quantum dynamics is mapped to an ensemble of trajectories. The mapping involves the measurements of the initial and final states. Therefore any possible coherence or entanglement is wiped out in the two-point measurement scheme. However, trajectories obtained from the global measurements may include the statistics about quantum correlation, such as the trajectory $\bar{\mathcal P}[\tau]$ defined in Eq. (\ref{eq:P'[tau]}).

As we know that the stochastic mutual information change $\Delta\iota$ is the mismatch between the joint stochastic entropy production $\sigma_{SR}$ of the system and the reference, and the stochastic entropy production of the system $\sigma_{S}$, namely
\begin{equation}
\Delta\iota = \sigma_{SR} - \sigma_{S}.
\end{equation}
Meanwhile, the mutual information change $\Delta\mathcal I$ (the average of the stochastic mutual information change) is also the mismatch between the average of the joint entropy production $\sigma_{SR}$ of the system and the reference, and the average of the entropy production of the system $\sigma_{S}$, namely
\begin{equation}
\Delta\mathcal I = \langle \sigma_{SR} \rangle_{\bar{\mathcal P}[\tau]} - \langle \sigma_S\rangle_{\mathcal P[\gamma]}.
\end{equation}
If we can construct a joint distribution, which marginalizes to $\bar{\mathcal P}[\tau]$ and $\mathcal P[\gamma]$ separately, then taking the average of the stochastic mutual information change $\Delta\iota$ over such distribution would automatically reach the average mutual information change $\Delta\mathcal I$. If the distribution is properly normalized, then we would also automatically get the integral fluctuation theorem of $\Delta\iota$.

The two-point measurement trajectory $\bar{\mathcal P}[\tau]$ is obtained from the projection on the system-reference eigenbasis, while the trajectory $\mathcal P[\gamma]$ is obtained from the projection on the local basis of the system. If the system-reference is an entangled state, then the joint system-reference eigenoperator does not commute with the eigenoperator of the local system. Then it would be impossible to assign a joint distribution over the global trajectory $\bar{\mathcal P}[\tau]$ and the local trajectory $\mathcal P[\gamma]$. For one evolution, such as $U_{SE}$ in our study, we can obtain different trajectories based on the different two-point measurements. And there are no unified classical descriptions for these different trajectories (measurement statistics), also known as the quantum contextuality \cite{budroniQuantumContextuality2022}.

To reconcile the noncommutative issue in the two-point measurement scheme, Micadei et al. proposed the conditional trajectory where the statistics of the local operator are conditioned on the global operator \cite{micadeiQuantumFluctuationTheorems2020}. Based on their idea, we can define a joint probability distribution including both the statistics of the local and the global operators, given by
\begin{equation}
\mathcal P_\text{joint}[\zeta] = p(a,b|k)p(s,r|l)|\langle km|U_{SE}|ln\rangle|^2 p_lp_n,
\end{equation}
with the stochastic variables $\zeta = \{s,r,l,n,a,b,k,m\}$. Recall that each variable is given by
\begin{multline}
    \rho_S = \sum_s p_s|s\rangle_S\langle s|, \quad \rho_R = \sum_r p_r|r\rangle_R\langle r|,\\
    \rho_E = \sum_n p_n|n\rangle_E\langle n|,\quad \rho_{SR} = \sum_l p_l|l\rangle_{SR}\langle l|.
\end{multline}
Here $\rho_S$, $\rho_R$, and $\rho_E$ are the initial state of the system, the reference, and the environment respectively. The joint initial state of the system and the reference is denoted as $\rho_{SR}$. The final states give the variables
\begin{multline}
    \rho'_S = \sum_a p'_a|a\rangle_S\langle a|, \quad \rho'_R = \sum_b p'_b|b\rangle_R\langle b|,\\
    \rho'_E = \sum_m p'_m|m\rangle_E\langle m|,\quad \rho'_{SR} = \sum_k p'_k|k\rangle_{SR}\langle k|.
\end{multline}
Recall that the global trajectory
\begin{equation}
\bar{\mathcal P}[\tau] = |\langle km|U_{SE}|ln\rangle|^2 p_lp_n,
\end{equation}
with the stochastic variables $\tau = \{l,n,k,m\}$, is considered when we study the entropy production of the combined system and reference. It can be obtained from the two-point measurement scheme on the joint eigenstates of $\rho_{SR}$ and $\rho'_{SR}$. When one studies the entropy production of the system alone, the local trajectory 
\begin{equation}
\mathcal P[\gamma] = |\langle am|U_{SE}|sn\rangle|^2 p_sp_n,
\end{equation}
with the stochastic variables $\gamma = \{s,n,a,m\}$, is considered. One can easily see that the global trajectory $\bar{\mathcal P}[\tau]$ can be obtained by marginalizing the local variables $s$, $r$, $a$, and $b$ on $\mathcal P_\text{joint}[\zeta]$, namely
\begin{equation}
\bar{\mathcal P}[\tau] = \sum_{s,r,a,b} \mathcal P_\text{joint}[\zeta].
\end{equation}
However, the local trajectory $\mathcal P[\gamma]$ can not be properly defined from the joint distribution $\mathcal P_\text{joint}[\zeta]$. Specifically, we have
\begin{equation}
\mathcal P[\gamma] \neq \sum_{l,r,k,b}\mathcal P_\text{joint}[\zeta].
\end{equation}
The joint distribution $\mathcal P_\text{joint}[\zeta]$ is designed from the global two-point measurement scheme. The local operator statistics are obtained after the global measurement is performed. Therefore the local operator statistics obtained from the joint distribution $\mathcal P_\text{joint}[\zeta]$ is not the same as the local operator statistics obtained from $\mathcal P[\gamma]$. As a consequence, we can only obtain the fluctuation theorem of the entropy production of the combined system and reference from the trajectory $\mathcal P_\text{joint}[\zeta]$. Mathematically, we have
\begin{equation}
\langle e^{-\sigma_{SR}}\rangle_{\mathcal P_\text{joint}[\zeta]} = 1,
\end{equation}
with the entropy production $\sigma_{SR} = \ln p_l + \ln p_n - \ln p'_{k} - \ln p'_{m}$. However, the fluctuation theorem of the entropy production of the local system can not be established from the trajectory $\mathcal P_\text{joint}[\zeta]$. In other words, one can verify that
\begin{equation}
\langle e^{-\sigma_{S}}\rangle_{\mathcal P_\text{joint}[\zeta]} \neq 1,
\end{equation}
with the entropy production $\sigma_S = \ln p_s + \ln p_n - \ln p'_{a} - \ln p'_{m}$. Since the fluctuation theorem of $\sigma_S$ is not properly established, one can also verify that
\begin{equation}
\langle e^{-\Delta\iota}\rangle_{\mathcal P_\text{joint}[\zeta]} \neq 1,
\end{equation}
with the stochastic mutual information change $\Delta \iota = \sigma_{SR} - \sigma_{S}$. Therefore the informational fluctuation theorem can not be established according to the joint conditional trajectory $\mathcal P_\text{joint}[\zeta]$. If the joint conditional trajectory $\mathcal P_\text{joint}[\zeta]$ is imposed in our tripartite setup, only the fluctuation theorem of the classical information dissipation can be obtained \cite{zhangConditionalEntropyProduction2021a}. 

On the contrary, the quasiprobability trajectory $\mathcal Q[\zeta]$ defined in Eq. (\ref{eq:Q_zeta}) correctly intertwines with the local state entropy production $\sigma_S$ and the global state entropy production $\sigma_{SR}$, as shown in Eq. (\ref{eq:sigma_s_sr}). It is impossible to construct a classical trajectory (described by the classical probability) that gives both the correct marginalized local and global trajectories. Therefore the quasiprobability trajectory is inevitable for quantum processes. After all the joint conditional trajectory $\mathcal P_\text{joint}[\zeta]$ is still classical. As shown in \cite{pashayanEstimatingOutcomeProbabilities2015}, quantum processes described by the classical trajectory can be efficiently simulated on the classical computers, which obviously does not include all the quantum processes. 

Quasiprobability trajectories have negative values. Therefore Jensen's inequality can not be applied (in order to derive the quantum data-processing inequality from the informational quantum fluctuation theorem). However, the significance of the fluctuation theorem is not to prove the positivity of its first-order average. The fluctuation theorem works as a generating function on all orders of the distributions. As stated in \cite{merhavStatisticalPropertiesEntropy2010}: ``When applying Jensen’s inequality, it is felt that a great deal of valuable information concerning the statistics of the entropy production is lost.'' The physical interpretation of the informational fluctuation theorem is given by its high-order statistics describing the information dissipation process, such as the fluctuation-dissipation relation, Eq. (\ref{eq:FD_theorem}). And the negative quasiprobability distribution does not forbid us to obtain information of the high-order statistics.

Moreover, we also want to emphasize the significance of the negative quasiprobability. Instead of a ``defect'' as a distribution, many studies have shown the advantage of negative quasiprobability in different contexts. For example, the negative quasiprobability samplings distinguish the classical and quantum computations \cite{pashayanEstimatingOutcomeProbabilities2015}. In quantum metrology, the quantum advantage stems from the negative quasiprobability distributions, which output larger Fisher information \cite{arvidsson-shukurQuantumAdvantagePostselected2020,lupu-gladsteinNegativeQuasiprobabilitiesEnhance2022}. Based on our results, we conjecture that the error correction for the information dissipation with a negative quasiprobability distribution needs to have additional requirements. However, it is beyond the scope of our current work.

\subsection{Initial and final states distributions}

The Bell-diagonal state (between the system qubit and the reference qubit) can be easily prepared on quantum computers, where two ancillary qubits are required \cite{pozzobomPreparingTunableBelldiagonal2019}. The corresponding quantum circuit is 
\begin{equation}
\Qcircuit @C=1em @R=1em {
|0\rangle_1 && \gate{R_y(\theta_1)} & \ctrl{2} & \qw & \qw & \qw & \qw \\
|0\rangle_2 && \gate{R_y(\theta_2)} & \qw & \ctrl{2} & \qw & \qw & \qw \\
|0\rangle_R && \qw & \targ & \qw & \gate{H} & \ctrl{1} & \qw  \\
|0\rangle_S && \qw & \qw & \targ & \qw & \targ & \qw 
}
\end{equation}
where qubits 1 and 2 are ancillary qubits. Here $R_y(\theta)$ is the rotation gate on $y$-axis and $H$ is the Hadamard gate \cite{nielsenQuantumComputationQuantum2010}. The CNOT operation applying on the ancillary qubit 1 (2) and the R (S) qubit gives the mixed state of R (S) qubit. The latter Hadamard gate and CNOT gate transform the product states to the Bell states. Measuring the RS qubits on Bell basis (which is also the eigenbasis of the RS qubits), we have the probabilities
\begin{align}
    p_{l=0} = &\cos^2\left(\frac {\theta_1} 2\right)\cos^2\left(\frac{\theta_2} 2\right), \nonumber \\
    p_{l=1} = &\sin^2\left(\frac {\theta_1} 2\right)\cos^2\left(\frac{\theta_2} 2\right),\nonumber \\
    p_{l=2} = &\cos^2\left(\frac {\theta_1} 2\right)\sin^2\left(\frac{\theta_2} 2\right), \nonumber \\
    p_{l=3} = &\sin^2\left(\frac {\theta_1} 2\right)\sin^2\left(\frac{\theta_2} 2\right).
\end{align}
Then local computational basis measurement or the Bell measurement on SR can reveal the initial distribution of $\rho_{SR}$. In our experiments, we randomly choose the initial distribution, where the angles are set as $\theta_1 = 0.7098\pi$ and $\theta_2 = 1.7059\pi$. It is easy to verify that the Bell-diagonal state generated by the above two angles are entangled.  

The thermal qubit (in the computational basis) can be prepared with one ancillary qubit by the circuit
\begin{equation}
\Qcircuit @C=1em @R=1em {
|0\rangle_E && \gate{R_y(\theta_3)} & \ctrl{1} & \qw \\
|0\rangle_3 && \qw & \targ &  \qw & \\
}
\end{equation}
where qubit 3 is ancillary. The angle $\theta_3$ is determined by the effective temperature $\beta$, namely $\theta_3 = 2\arctan(e^{\beta})$.

We set the unitary interaction between the system and environment qubits as the controlled gate $U_{SE} = |0\rangle_S\langle0|\otimes 1\!\!1_E+|1\rangle_S\langle1|\otimes Y_E$, which can be decomposed as one CNOT gate and one CZ gate. Although we can choose the interaction between the system and the environment qubits arbitrarily, choosing the specific $U_{SE}$ interaction is to take the advantage that the controlled $U_{SE}$ gate, given by $|0\rangle\langle0|\otimes 1\!\!1_{SE}+|1\rangle\langle 1|\otimes U_{SE}$, is the simplest three-qubit controlled gate to be realized on the IBM quantum computers \cite{maslovAdvantagesUsingRelativephase2016}. The controlled $U_{SE}$ gate is required to measure the quasiprobabilistic trajectory.

\begin{figure*}[t]
\centering
\includegraphics[width=\textwidth]{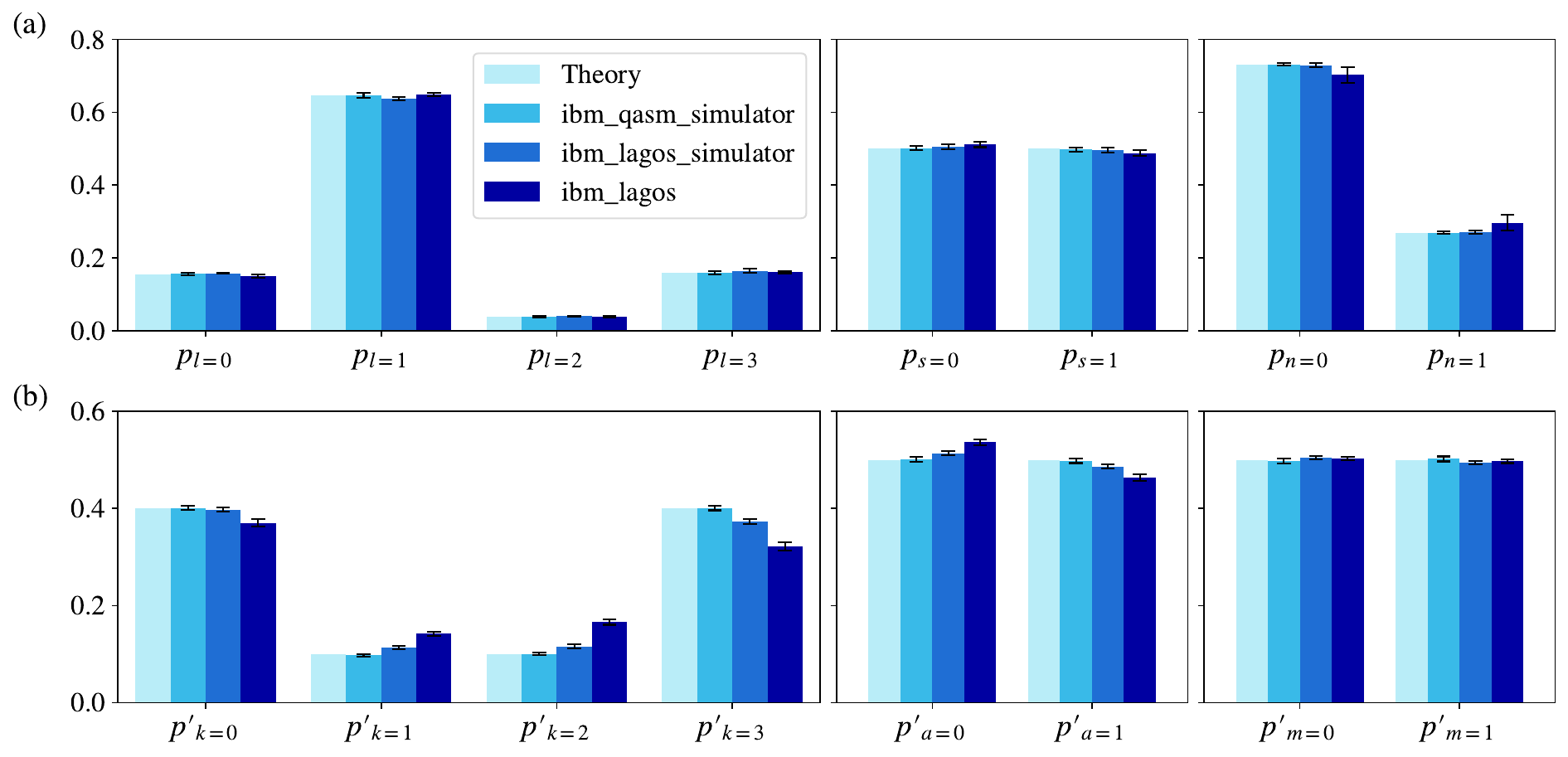}
\caption{Statistics of the initial and final states measured by the IBM quantum computer {\fontfamily{qcr}\selectfont ibm\_lagos}. (a) is for the initial state and (b) is for the final state. The {\fontfamily{qcr}\selectfont ibm\_qasm\_simulator} and {\fontfamily{qcr}\selectfont ibm\_lagos\_simulator} are provided by Qiskit, which classically simulates the quantum computers without and with the noises. The probabilities are estimated based on 8192 shots of the circuits. The error bars are given by $10\times 8192$ shots of the circuits. }
\label{fig_initial_final}
\end{figure*}

Combined with the initial state setup, the final state is given by the circuit 
\begin{equation}
\Qcircuit @C=1em @R=1em {
|0\rangle_{0} && \gate{R_y(\theta_1)} & \ctrl{2} & \qw & \qw & \qw & \qw & \qw & \qw \\
|0\rangle_{1} && \gate{R_y(\theta_2)} & \qw & \ctrl{2} & \qw & \qw & \qw & \qw & \qw \\
|0\rangle_R && \qw & \targ & \qw & \gate{H} & \ctrl{1} & \qw & \qw & \qw \\
|0\rangle_S && \qw & \qw & \targ & \qw & \targ & \ctrl{1} & \ctrl{1} & \qw \\
|0\rangle_E && \gate{R_y(\theta_3)}  & \ctrl{1}  & \qw & \qw & \qw  &\targ &  \control \qw & \qw  \\
|0\rangle_{3} && \qw & \targ & \qw & \qw & \qw & \qw & \qw & \qw
}
\end{equation}
Measuring the qubits R, S, and E in the computational basis can reveal the statistics of the final state. Note that the final state $\rho'_{SR}$ is no longer entangled (but classically correlated). The measured results, given by the simulators and the real quantum processor {\fontfamily{qcr}\selectfont ibm\_lagos} are shown in Fig. \ref{fig_initial_final}.

\subsection{Measuring the quasiprobabilistic trajectory}

The quasiprobabilistic trajectory $\mathcal Q[\zeta]$ given in Eq. (\ref{eq:example_Q}) is the multiplication of three amplitudes and the distribution of the initial state. The amplitude can be measured in a interference-based manner, reported in \cite{yungerhalpernJarzynskilikeEqualityOutoftimeordered2017}. Suppose that we want to measure the amplitude $\langle f|U|f\rangle$ in terms of the unitary evolution $U$. Firstly, we prepare the state $|\psi_{t_0}\rangle = \left(|0\rangle+|1\rangle\right)|f\rangle/\sqrt 2$, where the first qubit provides the supperposition for interference. Then perform the controlled $U$-gate where the first qubit is the control, which gives $|\psi_{t_1}\rangle = \left(|0\rangle|f\rangle+|1\rangle U|f\rangle\right)/\sqrt 2$. After that, the single-qubit gate $R_y(\theta)$ is applied to the first qubit, which gives $|\psi_{t_2}\rangle = \left(R_y(\theta)|0\rangle|f\rangle+R_y(\theta)|1\rangle U|f\rangle\right)/\sqrt 2$. Then the probability that the state $|\psi_{t_2}\rangle$ is projected on $|0\rangle|f\rangle$ is $P(0,f) = |\langle 0|\langle f|\psi_{t_2}\rangle|^2$, which is equal to
\begin{multline}
    \label{eq:p(0,f)}
    P(0,f) = \\
    \frac 1 2\left(\cos^2\left(\frac\theta 2\right)+\sin^2\left(\frac \theta 2\right)|\langle f|U|f\rangle|^2
    - \sin(\theta)\re \langle f|U|f\rangle\right).
\end{multline}
And the measured result $|1\rangle|f\rangle$ has the probability
\begin{multline}
    \label{eq:p(1,f)}
    P(1,f) = \\
    \frac 1 2\left(\sin^2\left(\frac\theta 2\right)+\cos^2\left(\frac \theta 2\right)|\langle f|U|f\rangle|^2
    + \sin(\theta)\re \langle f|U|f\rangle\right).
\end{multline}
Knowing the probabilities $P(0,f)$ (or $P(1,f)$) and $|\langle f|U|f\rangle|^2$, one can infer the amplitude $\re \langle f|U|f\rangle$. The imaginary part of $\langle f|U|f\rangle$ can be similarly measured by replacing the single-qubit gate $R_y(\theta)$ with $R_x(\theta)$.

Converting the above scheme to the quantum circuit model, we have
\begin{equation}
\label{qc:measure_amplitude}
\Qcircuit @C=1em @R=1.2em {
|0\rangle && \gate{H} & \ctrl{1} \qw & \gate{R_y(\theta)} & \meter \\
|0\rangle && \multigate{3}{V} & \multigate{3}{U} & \multigate{3}{V^\dag} & \meter \\
|0\rangle && \ghost{V} & \ghost{U} & \ghost{V^\dag} & \meter \\
\vdots &&  & & & \vdots\\
|0\rangle && \ghost{V} & \ghost{U} & \ghost{V^\dag} &  \meter
}
\end{equation}
The first qubit is the ancillary qubit providing the space for interference. Here $V$ is the operation preparing the state $|f\rangle$. If the state $|f\rangle$ is not in the computational basis, then the unitary operation $V^\dag$ is required in order to project on $|f\rangle$. If we need to measure the amplitude $\langle f|U|f'\rangle$, then additional unitary operations are required to transform $|f'\rangle$ to $|f\rangle$. In our example, measuring the amplitudes $\langle am|U_{SE}|sn\rangle$ and $\langle rs|\psi(l)\rangle$ needs three qubits, while measuring the amplitude $\langle \psi(l) n|U_{SE}|bam\rangle$ requires four qubits. 

\begin{figure*}[t]
\centering
\includegraphics[width=\textwidth]{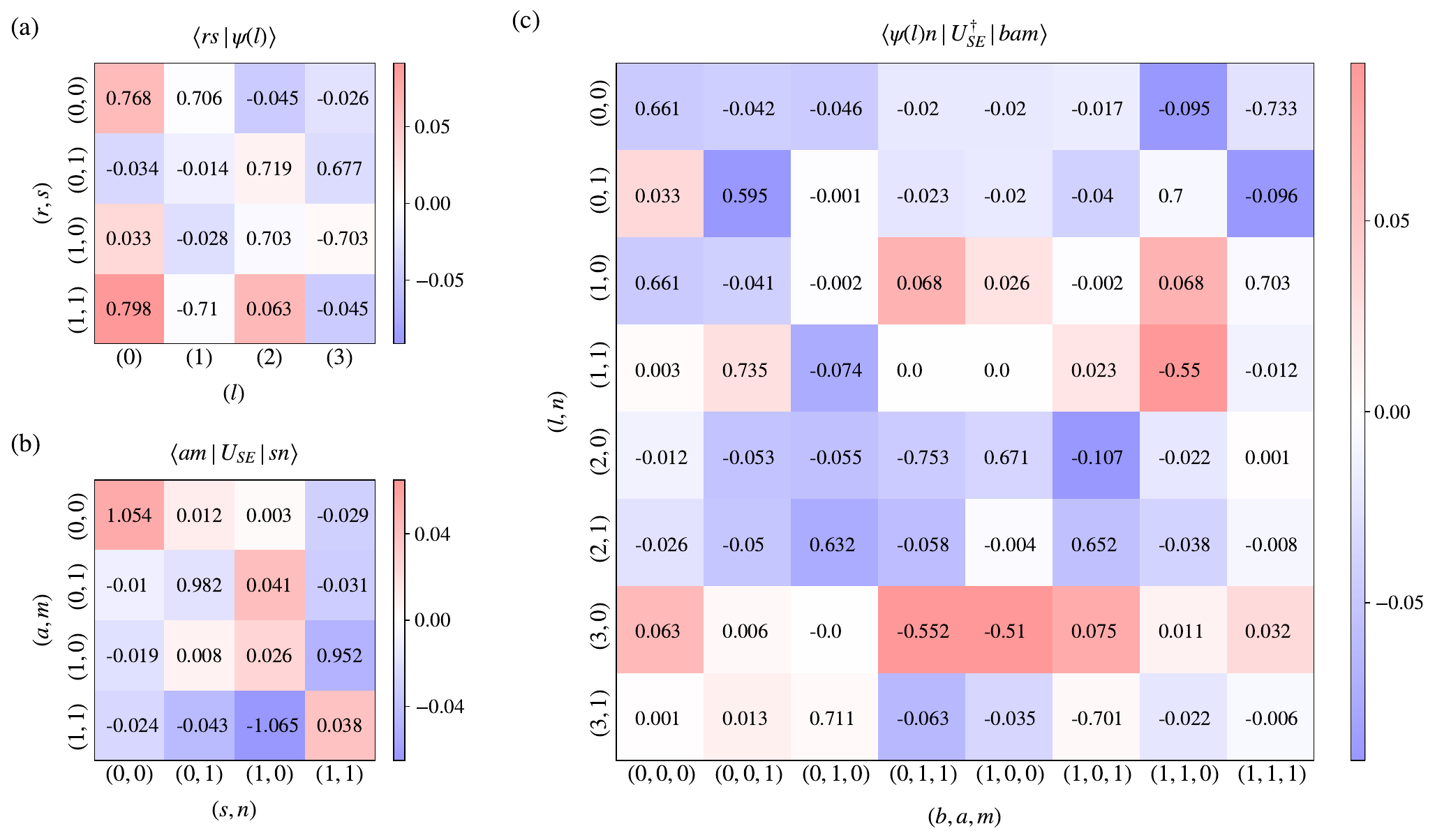}
\caption{Quasiprobability amplitudes measured on IBM quantum processor {\fontfamily{qcr}\selectfont ibm\_lagos}. Amplitudes $\langle rs|\psi(l)\rangle$, $\langle am|U_{SE}|sn\rangle$ and $\langle \psi(l) n|U^\dag_{SE}|bam\rangle$ are measured in (a), (b), and (c) respectively. The color represents the deviations from the theoretical values of $\langle rs|\psi(l)\rangle$, $\langle am|U_{SE}|sn\rangle$ and $\langle \psi(l) n|U^\dag_{SE}|bam\rangle$. }
\label{fig_amplitude}
\end{figure*}

There is a free parameter $\theta$ in the above scheme. Based on Eqs. (\ref{eq:p(0,f)}) and (\ref{eq:p(1,f)}), we can solve the amplitude $\re \langle f|U|f\rangle$ as the function of the angle $\theta$ and the probabilities $P(0,f)$ and $P(1,f)$ (assuming $\sin\theta\neq 0$), given by
\begin{equation}
\label{eq:re_amplitude}
    \re \langle f|U|f\rangle =\tan\left(\frac\theta 2\right) P(1,f) - \cot\left(\frac\theta 2\right) P(0,f) +\cot\theta.
\end{equation}
The state-of-art quantum computers are still very noisy. Therefore the measured probabilities $P(0,f)$ and $P(1,f)$ are likely deviated from the theoretical values. However, if the deviation can be predicted, then choosing the appropriate angles $\theta$ can still give the accurate amplitude $\re \langle f|U|f\rangle$. In other words, the noisy deviations in $P(0,f)$ and $P(1,f)$ are cancelled with each other in Eq. (\ref{eq:re_amplitude}). We exploit such error mitigation schemes in our experiments. The angles are decided by the noisy simulation results. Qiskit includes the noisy simulation custom for each quantum processor, where the noises are predicted based on the benchmarked metrics of quantum computers. A systematical study on the above error mitigation scheme for measuring the amplitudes will be addressed in the future. The measured amplitudes, as well as compared to the theoretical values, are presented in Fig. \ref{fig_amplitude}.

\subsection{Experimental setups}

We conduct all the experiments on the IBM quantum processor {\fontfamily{qcr}\selectfont ibm\_lagos}, through the access of IBM Quantum Researchers Program. The quantum computer {\fontfamily{qcr}\selectfont ibm\_lagos} is a seven-qubit quantum computer with the quantum volume 32. The typical metrics of {\fontfamily{qcr}\selectfont ibm\_lagos} are as follows: the average CNOT errors: $1.092\times 10^{-2}$; the average readout errors: $1.197\times 10^{-2}$; the average T1 time: 144.52 us; the average T2 time: 103.14 us. The experiments are completed in two months due to the limited monthly allocation of running hours. Therefore the metrics of {\fontfamily{qcr}\selectfont ibm\_lagos} changed in time. Both the circuits on measuring the initial or final state distribution and the quasiprobabilistic trajectories are submitted with $8192$ shots. Then the probabilities are estimated by the distribution of 8192 outputs. The standard deviations are obtained via repeating 10 times of the circuits with $8192$ shots. 

	
\section{Acknowledgments}

K.Z. was supported by the National Natural Science Foundation of China under Grant Nos. 12305028 and 12247103, and the Youth Innovation Team of Shaanxi Universities. We acknowledge the access to advanced services provided by the IBM Quantum Researchers Program. In this paper we used {\fontfamily{qcr}\selectfont ibm\_lagos}, which is one of the IBM Quantum Canary Processors. The views expressed are those of the authors, and do not reflect the official policy or position of IBM or the IBM Quantum team. The authors thank Vladimir Korepin, Xuanhua Wang, Qian Zeng, Wei Wu, Wufu Shi, He Wang and Hong Wang for helpful discussions.

\section{Competing interests}

The authors declare no competing financial or non-financial interests.

\section{Data availability}

The data that support the findings of this study are available from the corresponding author upon reasonable request.

\section{Code availability}

The Qiskit codes of this study are available from the corresponding author upon reasonable request.

\section{Author contributions}

K.Z. performed the analytical calculations, designed and performed the experiment. All authors contributed to writing the paper. The project was designed and supervised by J.W.



\providecommand{\noopsort}[1]{}\providecommand{\singleletter}[1]{#1}%

\end{document}